\documentclass{aa}
\usepackage{graphicx}

\def\gtrsim{\mathrel{\hbox{\rlap{\hbox{\lower4pt\hbox{$\sim$}}}\hbox{$>$}}}}
\def\ltsim{\mathrel{\hbox{\rlap{\hbox{\lower4pt\hbox{$\sim$}}}\hbox{$<$}}}}

\begin{document}

\title{A dominant magnetic dipole for the evolved Ap star candidate EK Eridani\thanks{Based on data obtained using the T\'elescope Bernard Lyot at Observatoire du Pic du Midi, CNRS/INSU and Universit\'e de Toulouse, France.}}

\author{M. Auri\`ere\inst{1,2}, R. Konstantinova-Antova\inst{3,1}, P. Petit\inst{1,2}, T. Roudier\inst{1,2},  J.F. Donati\inst{1,2},
C. Charbonnel\inst{4,2}, B. Dintrans\inst{1,2}, F. Ligni\`eres\inst{1,2}, G.A. Wade\inst{5}, A. Morgenthaler\inst{1,2}, S. Tsvetkova\inst{3} 
 }
\offprints{michel.auriere@ast.obs-mip.fr}
\institute{Universit\'e de Toulouse, UPS-OMP, Institut de Recherche en Astrophysique et Plan\'etologie,Toulouse, France\\
\email{[michel.auriere;pascal.petit;thierry.roudier;jean-francois.donati;boris.dintrans;francois.lignieres;\\
audrey.morgenthaler]irap.omp.eu}
\and
CNRS, Institut de Recherche en Astrophysique et Plan\'etologie, 14  Avenue Edouard Belin, 31400 Toulouse, France
\and
Institute of Astronomy, Bulgarian Academy of Sciences, 72 Tsarigradsko shose, 1784 Sofia, Bulgaria\\
\email{[renada;stsvetkova]@astro.bas.bg}
 \and
Geneva Observatory, University of Geneva, 51 Chemin des Maillettes, 1290 Versoix, Switzerland\\
\email{corinne.charbonnel@unige.ch}
 \and
Department of Physics, Royal Military College of Canada,
 PO Box 17000, Station 'Forces', Kingston, Ontario, Canada K7K 4B4\\
\email{gregg.wade@rmc.ca}
 }
 \date{Received ??; accepted ??}

\abstract
 {EK Eri is one of the most slowly rotating active giants known, and has been proposed to be the descendant of a strongly magnetic Ap star. }
{We have performed a spectropolarimetric study of EK Eri over 4 photometric periods with the aim of inferring the topology of its magnetic field.}
{We used the NARVAL spectropolarimeter at the Bernard Lyot telescope at the Pic du Midi Observatory, 
along with the least-squares deconvolution method, to extract high signal-to-noise ratio Stokes $V$ profiles from a timeseries of 28 polarisation spectra. We have derived the surface-averaged longitudinal magnetic field $B_{\ell}$. We fit the Stokes $V$ profiles with a model of the large-scale magnetic field and obtained Zeeman Doppler images of the surface magnetic strength and geometry.}
{$B_{\ell}$ variations of up to about 80 G are observed without any reversal of its sign, and which are in phase with photometric ephemeris. The activity indicators are shown to vary smoothly on a timescale compatible with the rotational period inferred from photometry ($308.8$~d.), however large deviations can occur from one rotation to another. The surface magnetic field variations of EK Eri appear to be dominated by a strong magnetic spot (of negative polarity) which is phased with the dark (cool) photometric spot. Our modeling shows that the large-scale magnetic field of EK Eri is strongly poloidal. For a rotational axis inclination of $i$ = 60$^\circ$, we obtain a model that is almost purely dipolar.}
{In the dipolar model, the strong magnetic/photometric spot corresponds to the negative pole of the dipole, which could be the remnant of that of an Ap star progenitor of EK Eri.  Our observations and modeling conceptually support this hypothesis, suggesting an explanation of the outstanding magnetic properties of EK Eri as the result of interaction between deep convection and the remnant of an Ap star magnetic dipole.
}

   \authorrunning {M. Auri\`ere et al.}
   \titlerunning {Topology of the magnetic field in EK Eri}

\maketitle

\section{Introduction}

The slowly rotating active G8 III-IV giant EK Eri (HR 1362, HD 27536) is  considered as a candidate for being the descendant of a strongly magnetic Ap star ( Stepie\'n 1993, Strassmeier et al. 1999).  The detection of a strong surface magnetic field, dominated by a poloidal component, is consistent with this hypothesis (Auri\`ere et al. 2008). EK Eri has now been surveyed photometrically during more than 30 years and its magnetic activity studied spectroscopically (Dall et al. 2010, hereafter DBSS10). These authors demonstrate that, while the photometric variation and possibly the period may change from season to season, an average period of 308 $\pm$ 2.5 d provides an acceptable explanation of the data. DBSS10 report a photometric ephemeris based on this period, confirm that the main light variations are due to dark spots and tentatively suggest that the star is a dipole-dominated oblique rotator viewed close to equator-on. Based on this picture they propose that the rotation period is twice the photometric period. 

Following the initial magnetic observations presented by Auri\`ere et al. (2008), we have now monitored EK Eri throughout 4 seasons, corresponding to about 4 photometric periods ($308.8$~d., according to DBSS10). In this paper we present our new data and the associated analysis, which enables us to establish the rotational period of EK Eri and to model its surface magnetic topology.

\section{Observations}

\subsection{Observations with NARVAL}

The 28 observations of EK Eri were obtained between 20 September 2007 and 21 March 2011 at the 2m telescope Bernard Lyot (TBL) of Pic du Midi Observatory. We used the NARVAL spectropolarimeter, which is a copy of the 
ESPaDOnS instrument at the Canada-France-Hawaii Telescope (Donati et al. 2006a). NARVAL was used in polarimetric mode with a spectral resolution of 
about 65000, in the same configuration and with the same procedure as described by Auri\`ere et al. (2008).  

To complete the Zeeman analysis, least-squares deconvolution (LSD, Donati et al. 1997) was applied to all observations. We used a line mask calculated for an effective temperature of $5250$~K, $\log g =3.5$, and a microturbulence of 2.0 km s$^{-1}$, consistent with physical parameters reported by DBSS10. In the present case this method enabled us to average about 11,000 spectral lines and to derive Stokes $V$ profiles with a S/N that improved by a factor of about 40 in comparison with that for single lines.  We then computed the longitudinal magnetic field $B_{\ell}$ in Gauss, using the first-order moment method (Rees \& Semel 1979, Donati et al. 1997). 

Spectra obtained with NARVAL enable us to simultaneously obtain magnetic data, measurements of activity indicators as well as radial velocity ($RV$) measurements, so that studying correlations between these quantities is straightforward. The activity of the star was monitored by using line--activity indicators measured from the NARVAL spectra. We computed the $S$-index (defined from the Mount Wilson survey, Duncan et al. 1991) for the chromospheric Ca~{\sc ii} H \& K line cores. Our procedure was first calibrated on the main sequence solar-type stars of Wright et al. (2004), then we added 0.03 to the index to fit measurements of 5 giant stars observed by Duncan et al. (1991) and Young et al. (1989). We also measured the relative intensity with respect to the continuum ($R_{\rm c}$) of the Ca~{\sc ii} IR triplet (854.2 nm component) and  H$\alpha$. 

We measured the $RV$ of EK Eri from the LSD Stokes $I$ profiles using a gaussian fit.

Table 1 reports the date, HJD, photometric phase (based on the ephemeris of DBSS10) , the Ca~{\sc ii} H \& K  $S$-index, $R_{\rm c}$ values for Ca~{\sc ii} 854.2 nm and H$\alpha$, the measured $B_{\ell}$ values and their uncertainties, and the inferred radial velocities.

\begin{table*}
\caption{Log of observations of EK Eri (for details, see Sect. 2.1)}           
\centering                         
\begin{tabular}{c c c c c c c c c }     
\hline\hline               
Date      & HJD      &Phot.&$S$-index& Ca~{\sc ii} &H$\alpha$& B$_l$& $\sigma$& $RV$  \\
          &2450000+  &Phase&         & 854.2 nm   &         & G    &  G      & km s$^{-1}$   \\
\hline                        
20 Sep. 07 & 4364.69  &0.212& 0.501 & 0.319       & 0.226   &-98.6 & 1.0     & 7.118  \\
12 Nov. 07 & 4417.44  &0.383& 0.457 & 0.291       & 0.233   &-21.1 & 0.7     & 7.047  \\
13 Nov. 07 & 4418.45  &0.386& 0.456 & 0.297       & 0.233   &-21.4 & 0.9     & 7.053  \\
19 Jan. 08 & 4485.31  &0.603& 0.470 & 0.303       & 0.230   &-28.9 & 1.2     & 7.096  \\
20 Jan. 08 & 4486.38  &0.606& 0.463& 0.295       & 0.227   &-31.6 & 0.9     & 7.110  \\
06 Feb. 08 & 4503.35  &0.661& 0.469& 0.302       & 0.234   &-45.6 & 0.8     & 7.111  \\
03 Apr. 08 & 4560.31  &0.845& 0.513& 0.339       & 0.250   &-62.7 & 1.2     & 7.145  \\
28 Aug. 08 & 4707.68  &0.323& 0.502& 0.331       & 0.251   &-12.8 & 0.7     & 7.052  \\
14 Sep. 08 & 4724.60  &0.378& 0.511& 0.338       & 0.262   &-27.9 &1.4      & 7.043  \\
24 Sep. 08 & 4734.57  &0.410& 0.510& 0.341       & 0.270   &-18.8 &1.1      & 7.000  \\
29 Sep. 08 & 4739.71  &0.426& 0.506& 0.338       & 0.264   &-13.7 &0.8      & 6.996  \\
20 Dec. 08 & 4821.37  &0.691& 0.543& 0.365       & 0.266   &-36.8 &1.1      & 7.046  \\
30 Jan. 09 & 4862.35  &0.824& 0.588& 0.390       & 0.291   &-46.4 &1.0      & 7.207  \\
25 Feb. 09 & 4888.31  &0.908& 0.573& 0.369       & 0.270   &-67.8 &0.9      & 7.176  \\
28 Sep. 09 & 5103.62  &0.605& 0.602& 0.423       & 0.286   &-48.0 &1.2      & 7.080  \\
27 Oct. 09 & 5132.57  &0.699& 0.576& 0.396       & 0.280   &-56.9 &1.3      & 7.101  \\
15 Jan. 10 & 5212.44  &0.957& 0.547& 0.339       & 0.254   &-88.6 &2.7      & 7.190  \\
03 Feb. 10 & 5231.32  &0.018& 0.566& 0.354       & 0.259   &-90.7 &2.1      & 7.200  \\
13 Feb. 10 & 5241.29  &0.051& 0.575& 0.363       & 0.263   &-90.9 &2.1      & 7.214  \\
06 Mar. 10 & 5262.30  &0.119& 0.548& 0.353       & 0.258   &-77.0 &1.2      & 7.218  \\
22 Mar. 10 & 5278.32  &0.171& 0.533& 0.348       & 0.260   &-58.9 &1.0      & 7.195  \\
21 Sep. 10 & 5461.62  &0.764& 0.598& 0.381       & 0.278   &-90.9 &1.2      & 7.062  \\
13 Oct. 10 & 5483.64  &0.836& 0.565& 0.370       & 0.262   &-94.6 &0.9      & 7.107  \\
12 Nov. 10 & 5513.55  &0.932& 0.571& 0.365       & 0.264   &-93.1 &1.1      & 7.183 \\
04 Dec. 10 & 5535.36  &0.003& 0.576& 0.359       & 0.270   &-87.2 &6.8      & 7.150 \\
03 Jan. 11 & 5565.28  &0.100& 0.585& 0.374       & 0.268   &-77.7 &1.0      & 7.150 \\
23 Jan. 11 & 5585.44  &0.165& 0.589& 0.436       & 0.272   &-57.6 &1.0      & 7.166 \\
21 Mar. 11 & 5642.31  &0.350& 0.574& 0.403       & 0.277   &-39.5 &1.2      & 7.130 \\
\hline  
\end{tabular}

Note: Date of mid-observation, corresponding HJD, photometric phase, $S$-index for Ca~{\sc ii} H \& K, $R_{\rm{c}}$ values for Ca~{\sc ii} 854.2 nm and H$\alpha$, $B_{\ell}$ values and their uncertainties, and radial velocity of the LSD Stokes $I$ profile.                            
\end{table*}

\begin{figure}
\centering
\includegraphics[width=9 cm,angle=0] {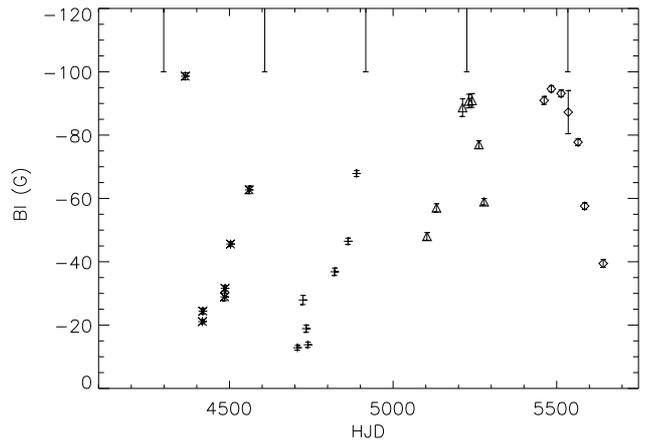} 

\caption{Variations of the longitudinal magnetic field ${B_\ell} $ in G versus HJD (minus 2450000). Different symbols represent different observing seasons: stars: 2007-2008, crosses: 2008-2009, triangles: 2009-2010, diamonds: 2010-2011. Vertical marks at the top of the figure indicate the times of photometric minima according to the ephemeris of DBSS10.}
\end{figure}

\begin{figure}
\centering
\includegraphics[width=8. cm,angle=0] {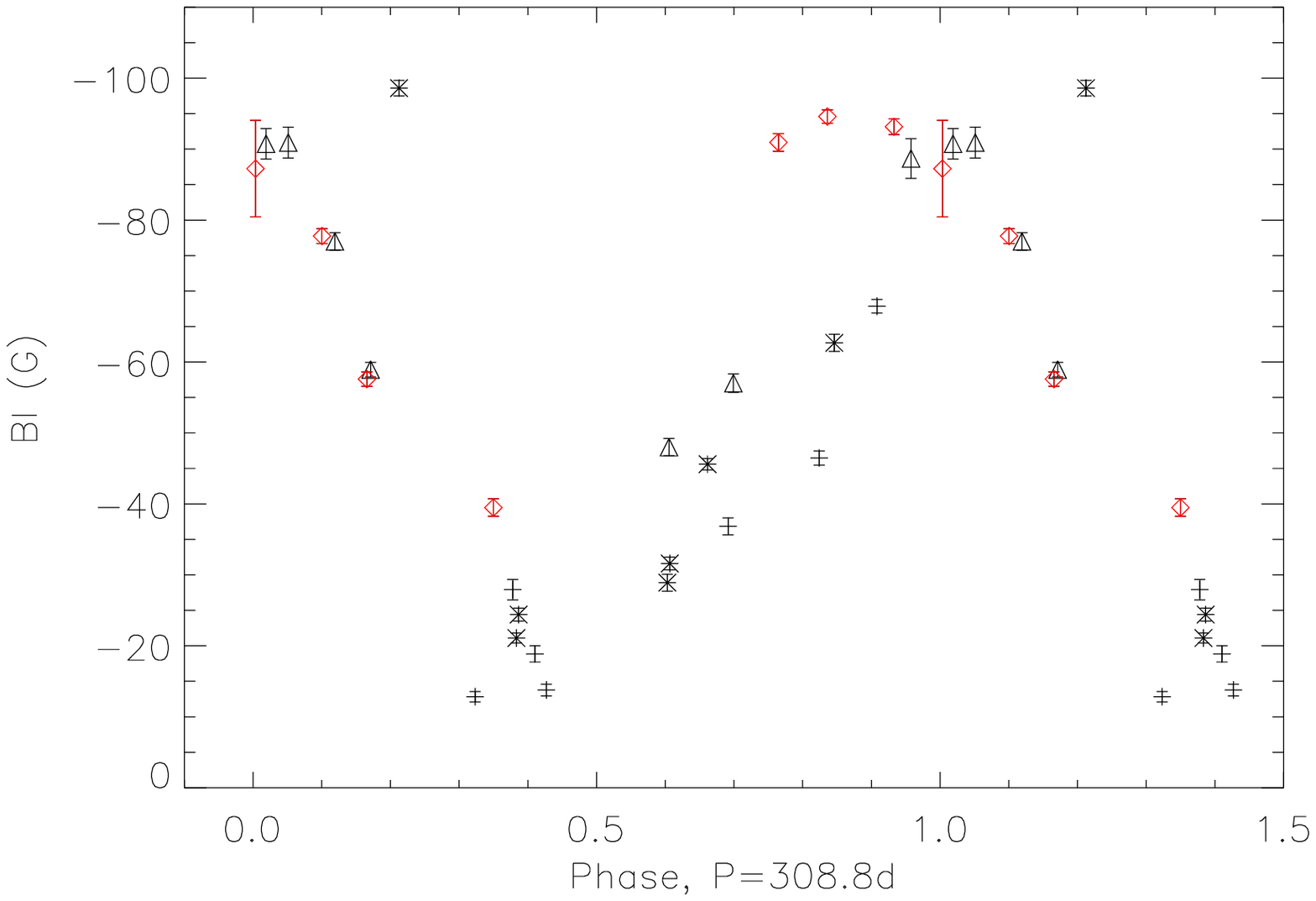} 
\includegraphics[width=8. cm,angle=0] {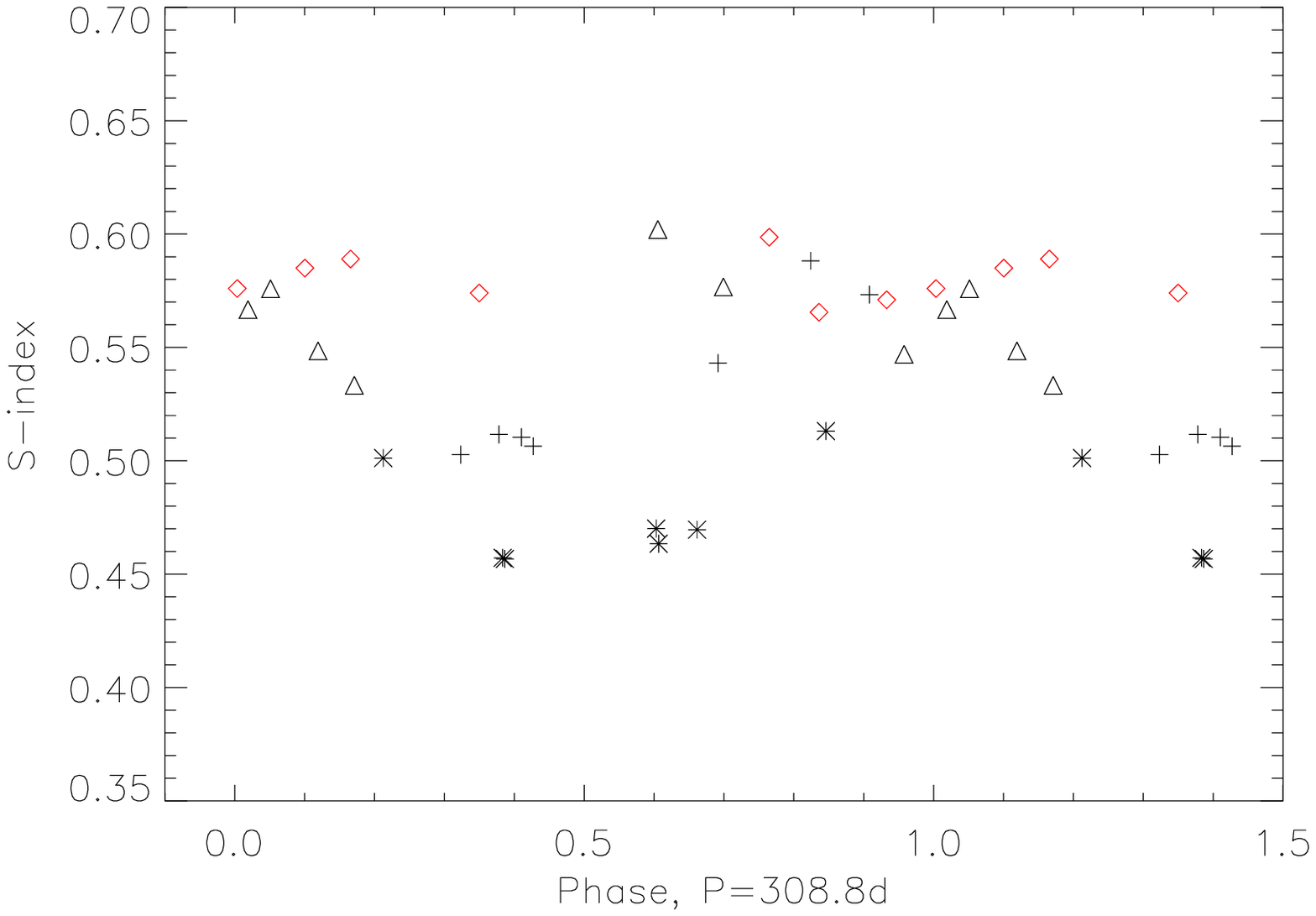}
\includegraphics[width=8. cm,angle=0] {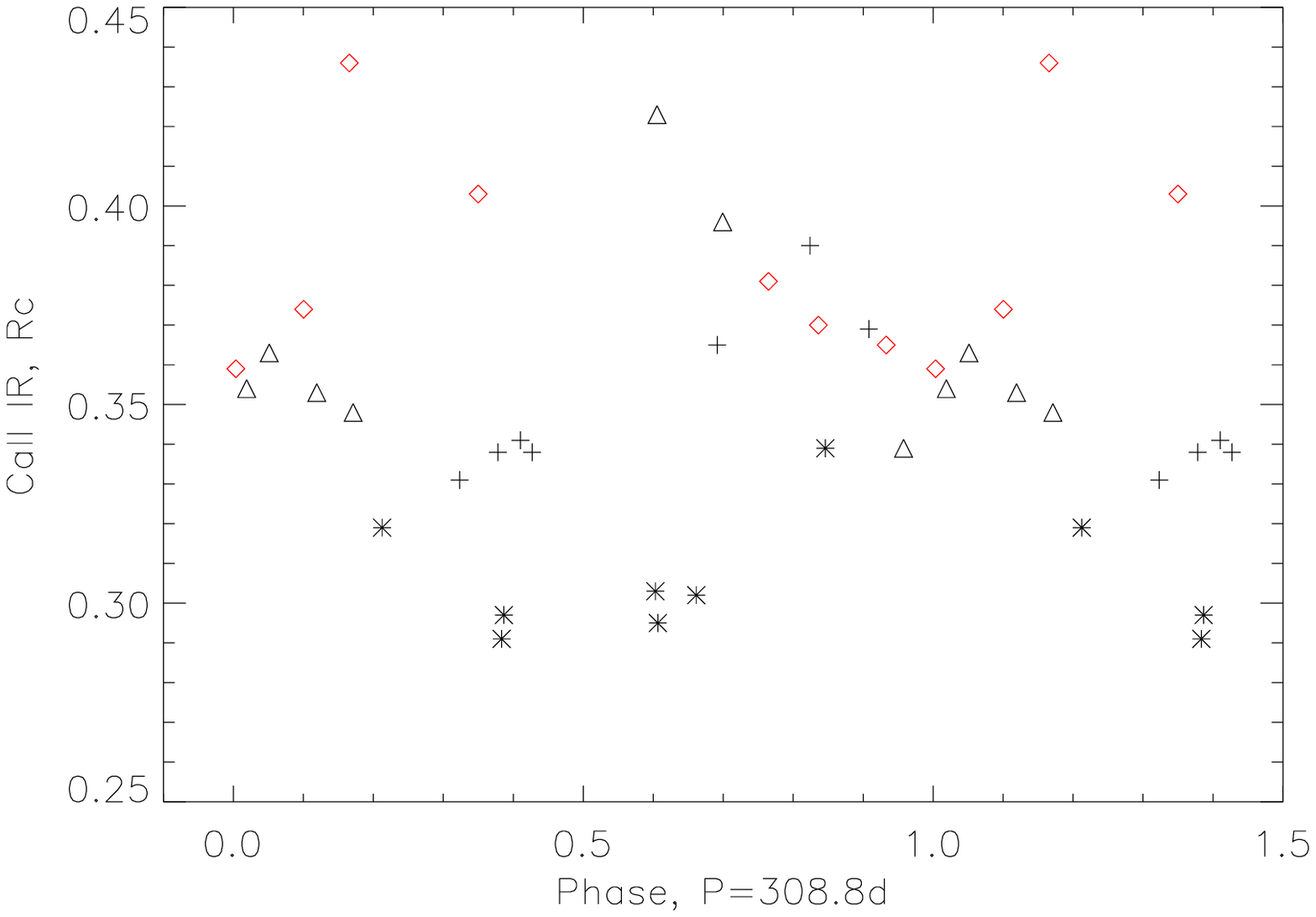}
\includegraphics[width=8. cm,angle=0] {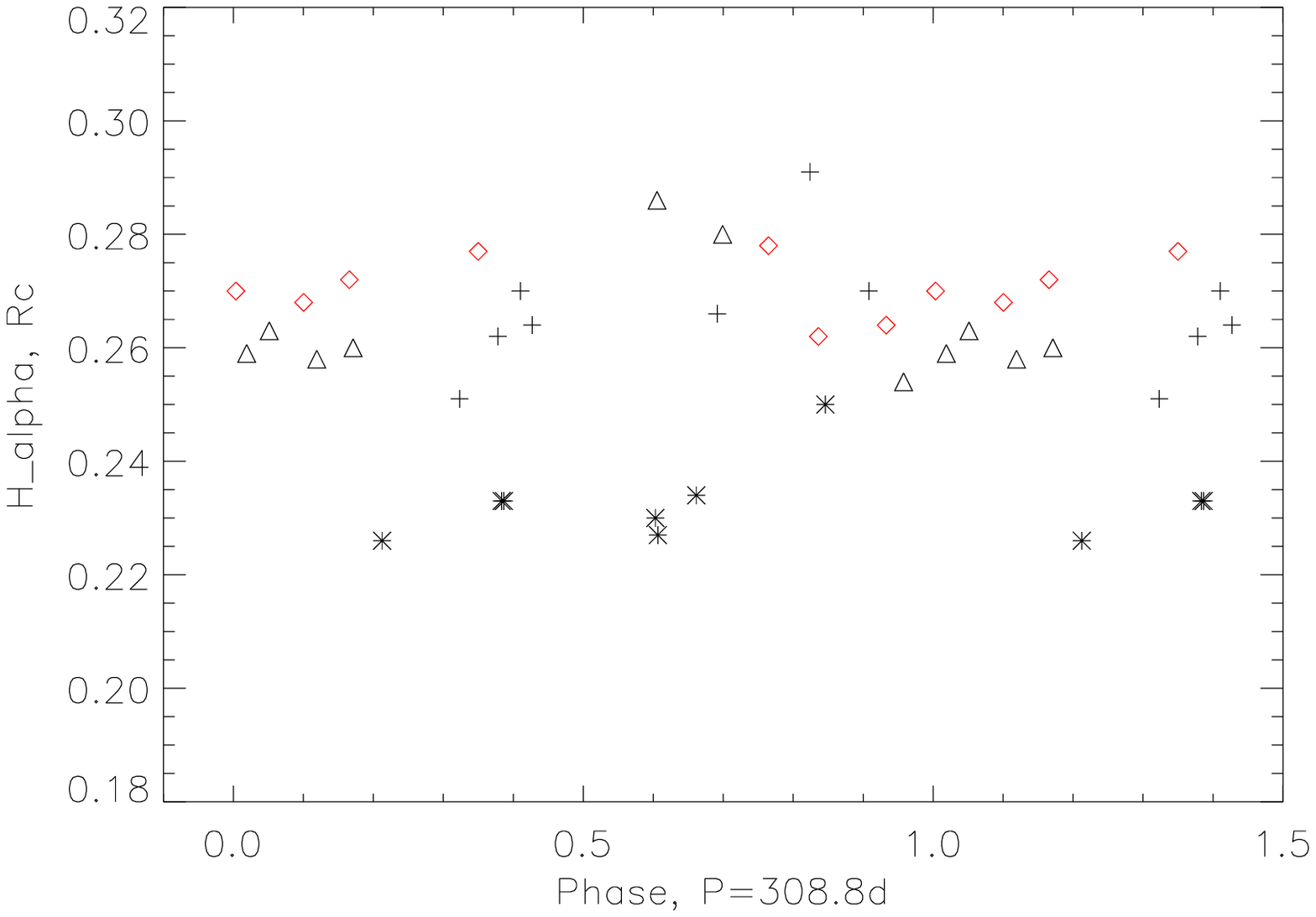}

\caption{Variations of the longitudinal magnetic field $B_{\ell}$ (upper panel) and of the activity indicators (lower panels), with photometric phase. The lower panels, from top to bottom, show:  $S$-index, $R_{\rm{c}}$ for Ca~{\sc ii} IR lines and H$_{\alpha}$. Different symbols represent different observing seasons: stars: 2007-2008, crosses: 2008-2009, triangles: 2009-2010, diamonds: 2010-2011; the 2010-2011 season is shown in red.} 
\end{figure}

\subsection{Results: variations of $B_{\ell}$}

Table 1 and Fig. 1 show that $B_\ell$ does not reverse its polarity during the 4 observed seasons. Figure 1 displays $B_{\ell}$ measurements with respect to HJD. Vertical marks at the top of Fig. 1 indicate the times of photometric minima according to the ephemeris of DBSS10, i.e. when the cool spot is centred on the visible hemisphere of the star (phase 0.0). We see that the unsigned $B_{\ell}$ is approximately at maximum at these dates. In the third season (2009-2010), which is well-centred on the spot passage, $|B_{\ell}|$ is confirmed to peak at phase 0. This trend is clearly visible in Fig. 2 (upper panel) which contains the same data as in Fig.1, but folded with respect to the photometric ephemeris (DBSS10, $P = 308.8$ days, HJD$_0$ = 245 3372.679). This is consistent with a picture in which the center of the dark spot of EK Eri coincides with the strongest $|B_{\ell}|$. DBSS10 suggested that the dark spot could correspond to the pole of the remnant of the dipole of an evolved Ap star. 

Since $B_{\ell}$ does not reverse its polarity, in this scenario, only one pole of the putative dipole is observed, and the rotational period would then be equal to the photometric period, and not its double, contrary to the suggestion of DBSS10. 

Figure 2 also shows that the (year to year) scatter of the points is significant with respect to the error bars. It could be due to an incorrect period, or in changes in activity level, as observed in photometry. In the scenario where the rotational period is the photometric one, this would show that an oblique rigid rotator model (ORM, Stibbs 1950) alone could not explain the variations of $B_{\ell}$. Some dynamo process would therefore need to be invoked.

\subsection{Results: variations of activity indicators and $RV$}

Figure 2 (lower panels)  shows the variations of the activity indicators versus photometric phase. Though the points are scattered, their lower envelope follows the variations of $|B_{\ell}|$ with respect to phase, with a maximum near phase 0, and a minimum near phase 0.5. This trend is more visible for Ca~{\sc ii} ($S$-index and IR lines) than for H$\alpha$. Some points are significantly above the lower contours of the plots and they are discussed below. In particular, during the  2010-2011 season, the modulation of activity indicators appears to disappear (this season is highlighted in red in Fig. 2). Error bars are computed for each measurement of Ca~{\sc ii} IR and H$\alpha$ measurement. They are generally of order 0.001, and never reach 0.01. As to $S$-index, an internal error of 0.007 is estimated for the whole sample (see also Sect. 2.4) in agreement with the day-to-day scatter (Table 1, 12/13 November 2007, 19/20 January 2008). These error bars for activity indicators are smaller than the symbols used in Fig. 2 (lower panel) and are therefore not plotted there.

 Figure 3 shows the variations of $RV$ versus phase. The long term stability of NARVAL is about 30 m s$^{-1}$ (e.g. Auri\`ere et al. 2009). $RV$ is found to vary with photometric phase, as already reported by Dall et al. (2005, DBSS10). There is some scatter near phase 0 and a delay of the $RV$ maximum by about 0.1 cycles with respect to $B_{\ell}$. 
This trend strongly suggests that the cool/magnetic spot is responsible for the $RV$ variations (see also Dall et al. 2005, DBSS10). Because of the small $v\sin i$ of EK Eri, classical Doppler imaging is not possible for this star. On the other hand, a small $v\sin i$ does not prevent mapping using Zeeman-Doppler imaging as described in Sect. 3. (e.g. Petit et al. 2008). 

\subsection{Discrepant points in  $B_{l}$ and activity indicators}

 Figure 2 shows that the seasonal variations of $B_{\ell}$ and the activity indicators generally vary coherently with photometric phase, and that there is seasonal scatter greater than the measurement error bars. In addition there are some outstanding deviations from the mean variation.

 As far as the $B_{\ell}$ measurements are concerned, the observations of 20 September 2007 (the first one, which is our ``discovery observation''), as well as those obtained on 21 September and 13 October 2010 present outstanding deviations from the general phase variations. 
On 20 September 2007, the $S$-index and other activity indicators do not deviate from the general phase variation. On the other hand, for the 2 first autumn 2010 observations, both $B_{\ell}$ and activity indicators are stronger than expected at their respective phases.

 Remarkably, in the 2010-2011 season, the modulation of activity indicators disappears: this season is highlighted in red in Fig. 2. While $B_{\ell}$ values approach the typical level and modulation at the end of 2010, the intensity of the activity indicators remains high.
 Also, on some occasions, in particular 28 September 2009, the activity indicators are surprisingly high, while the $B_{\ell}$ value is quite normal with respect to the phase. Looking closely at the data, we find that we may have a problem of normalization of the continuum for this date, which can affect the measurements of activity indicators, but not  $B_{\ell}$ (as the Stokes $V$ measurement is a differential measurement). To further investigate this potential background problem, we have computed the Ca~{\sc ii} H emission index used by Morgenthaler et al. (2011, 2012). This method makes use of  a synthetic spectrum from the POLLUX database (Palacios et al. 2010) to normalise the continuum. It was found to be very effective in the case of solar-type dwarfs, enabling those authors to reach internal errors of about 0.001 for chromospheric emission index measurements (Morgenthaler et al. 2011, 2012). However, in the case of EK Eri this method did not allow us to reduce the deviations, and we present our $S$-index measurements in Table 1 and Fig. 2 as they are.

 Of peculiar interest are the isolated enhancements of  $B_{\ell}$, activity indicators or both. They may be associated with flares as observed in active giants (e.g. Konstantinova-Antova et al. 2000, 2005). In this case it would be the first time that simultaneous observations of magnetic field and activity indicators have been performed during flares in the stellar context. In the solar case, a simultaneous increase of magnetic field and H$\alpha$ has been observed (Lozitsky et al. 2000).

 The outlying magnetic observations of September 2007 and autumn 2010 were found to increase dramatically the reduced $\chi^2$ of our model fitting in Sect. 3. We will use only data between November 2007 and March 2010 (spanning about 3 rotations) for forthcoming magnetic analysis.

\begin{figure}
\centering
\includegraphics[width=9 cm,angle=0] {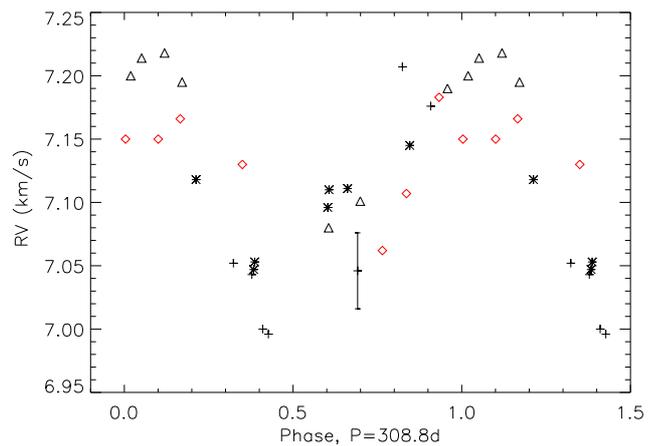}
\caption{Variations of radial velocity with photometric phase. An error bar of 30 m $s^{-1}$ is indicated. Different symbols represent different observing seasons: stars: 2007-2008, crosses: 2008-2009, triangles: 2009-2010, diamonds: 2010-2011; the 2010-2011 season is shown in red.}
\end{figure}

\section{The topology of the magnetic field at the surface of EK Eri}

\subsection {Zeeman-Doppler imaging}

In order to model our series of Stokes $V$ line profiles and to reconstruct the surface magnetic geometry of the star, we have used the Zeeman-Doppler imaging inversion method (ZDI; Donati and Brown  1997). The version of the code employed is that described by Donati et al. (2006b): the surface magnetic field is projected onto a spherical harmonics frame and the magnetic field is resolved into poloidal and toroidal components.
Because of the very slow rotation of EK Eri, its $v\sin i$ has been impossible to determine up to now due to the degeneracy between $vsini$ and macroturbulence (DBSS10). These authors consider that 0.8 km s$^{-1}$ is a safe upper limit for $v\sin i$. In our models we used $v\sin i$ = 0.5~km s$^{-1}$ and found that varying this value did not significantly change our results.
 We used a linear limb darkening coefficient equal to 0.75. We limited the spherical harmonics
expansion to $\ell \le 3$ since increasing this threshold do not significantly change the results.

\subsection {The rotational period of EK Eri}

Knowing the rotational period of EK Eri is essential to be able to perform ZDI. Photometric variations of EK Eri have been now monitored during more than 30 years and a period of 308.8 $\pm$ 2.5 days has been adopted (DBSS10) as well as an ephemeris. This period may have changed slightly during the span of the photometric observations (Strassmeier et al. 1999). DBSS10 argued that the rotational period could be equal to twice the photometric period, i.e. 617.6 $\pm$ 5 days.
To determine independently the rotational period of EK Eri, we have followed the approach of Petit et al. (2002), and calculated a set of magnetic maps, assuming for each map a different value for the rotational period. We impose a constant entropy for all the images and calculate as a goodness-of-fit parameter the reduced $\chi^2$ ($\chi^2 _r$ hereafter) by comparing the set of synthetic Stokes $V$ profiles produced by ZDI to the observed time-series of Stokes $V$ profiles. We then study the variations of $\chi^2 _r$ over the range of rotation periods, to determine the period value producing the best magnetic model (identified by the lowest value of $\chi^2 _r$). Here we scanned 400 values of the period between 200 and 700 days, a range which encompasses the photometric and rotational periods proposed for EK Eri by DBSS10. Our periodograms show a favoured period near 311 days. However the minimum $\chi^2 _r$ value is high (about 25), notwithstanding that we limited our observational sample to the timespan November 2007 - March 2010 in order to avoid the effect of the strong deviations observed on 20 September 2007 and during the autumn of 2010. We therefore introduced a possible differential rotation (Petit et al. 2002). Ultimately, we could not improve the results and deduced a negligible differential rotation for EK Eri. We also considered that our approximate line profile model might be the source of the high $\chi^2_r$: the current local line profile used in our model is a gaussian function. In an attempt to improve the fit of the line profile, we introduced a Lorentzian function and some asymmetry, but we still could not reduce significantly the $\chi^2 _r$. Ultimately, we consider that the relatively poor detailed fit to the Stokes $V$ profiles resulting from our best-fit model is mainly due to the seasonal scatter of the magnetic field strength. 

Figure 4 shows the periodogram obtained for the whole data set between November 2007 and March 2010, assuming a rotational axis inclination of EK Eri of $i = 60^\circ$ and truncating the expansion at $\ell=3$. 
For rotational periods of 311.5 d, 308.8 d, and 625 d, we obtain respectively values for $\chi^2 _r$ of 25 (our minimum), 25.3 and 38.2. The 600 day period suggested by DBSS10 is therefore eliminated both by our periodogram analysis and by the fact that their scenario to explain $P_{\rm rot} = 2P_{phot}$ is not supported by the $B_{\ell}$ variations (Sect. 2.2). Since we found that phase 0.0 of the ephemeris of DBSS10  corresponds to approximately the center of the magnetic spot which dominates the magnetic map (see below), we decided to use the photometric period of 308.8 days (and ${\rm HJD_0}$ = 245 3372.679) for our analysis of the magnetic topology.

Figure 5 presents the fit of our best model to the observed Stokes $V$ LSD profiles, for $P_{\rm rot}$ = 308.8 d, and $i = 60^\circ$.

\begin{figure}
\centering
\includegraphics[width=8 cm,angle=0] {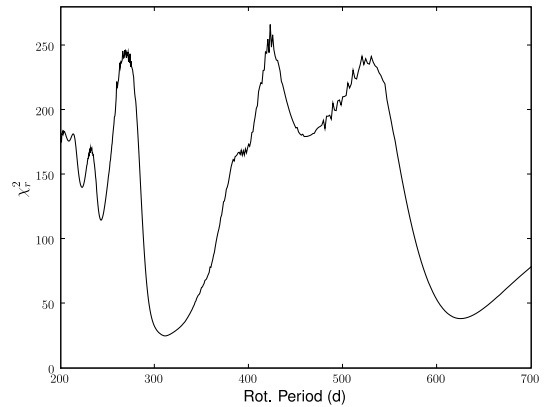} 
\caption{Periodogram obtained using the entire data set between November 2007 and March 2010.}
\end{figure}  

\begin{figure}
\centering
\includegraphics[width=6 cm,angle=0] {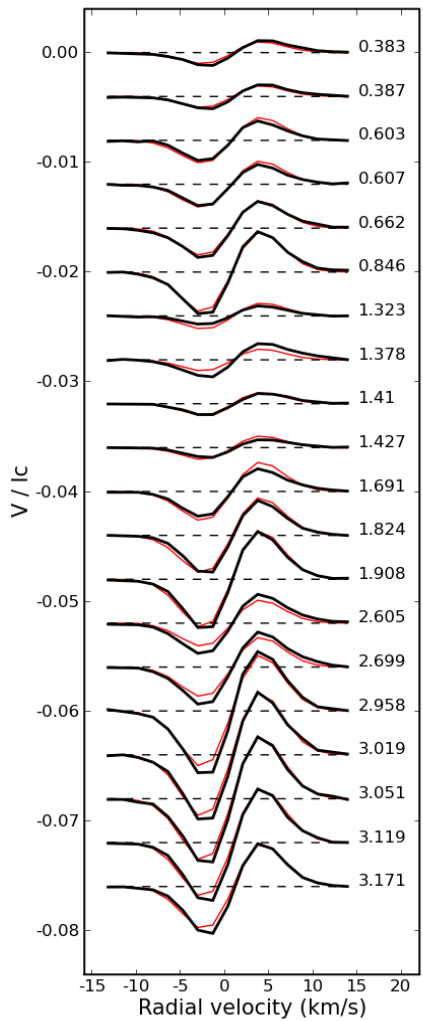} 
\caption{Fit of our best-fit model (for $P_{\rm rot}=308.8$~d, $i=60^\circ$) (thin red curves) to the Stokes $V$ LSD profiles (thick black curves), after correction of the mean radial velocity of the star. Rotational phases (according to the ephemeris of DBSS10) are indicated.}
\end{figure}

\subsection {Possible topologies of the surface magnetic field of EK Eri}

In the two previous subsections, we have presented the models and the parameters employed to infer the magnetic topology of EK Eri. Because $v\sin i$ cannot be determined precisely, the value of the inclination $i$ is also only weakly constrained. This is our main remaining uncertain parameter which we have attempted to constrain as part of the ZDI reconstruction. We have therefore fit our Stokes $V$ data with our models and $i$ varying from 85$^\circ$ to 40$^\circ$.  

For all models we find that the poloidal component contains more than 98$\%$ of the reconstructed magnetic energy. When $i$ decreases, we find  the ZDI procedure to provide us with a better magnetic model, in the sense that the total information content of the reconstructed magnetic geometry decreases (we follow here the maximum entropy criterion, Donati 2001, explicited in Morin et al. 2008). For high inclination, $i > 80^\circ$, the topology is dominated by a quadrupolar configuration. In this case, the two rotation poles correspond to positive polarity while there is a nearly equatorial band, with a single enormous magnetic spot, both of negative polarity. The magnetic spot corresponds to the phase of the dark (and therefore presumably cool) photometric spot to better than 0.1 cycles. 

For $i < 80^\circ$, the dipole component dominates and corresponds to about $90\%$ of the magnetic energy at about $i$ = 60$^\circ$. Figure 6 shows the corresponding ZDI map. This map also features a strong magnetic spot of negative polarity, corresponding to the negative pole of the magnetic dipole. Again the magnetic maximum corresponds in phase to the photometric spot. The characteristics of this model are compatible with the scenario of Auri\`ere et al. (2008) and DBSS10, in which the magnetic/photometric spot would correspond to the remnant of a magnetic pole of the Ap star progenitor of EK Eri. 

 Given the lower information content (i.e. higher entropy) associated with the predominantly dipolar configuration, we consider this model to be better, or at least more likely, than the quadrupolar model. In addition, this model suggests an obvious physical interpretation: that the corresponding magnetic pole is at the position of the photometric spot of EK Eri, and that the dipole could be the remnant of the magnetic field of an Ap star progenitor. Both Fig. 5 - presenting the variations of the Stokes $V$ profiles - and Fig. 2 (upper panel) - presenting the variations of $B_{\ell}$ - show that the magnetic field varies on a large scale, and does not experience a sudden drop when the ``magnetic spot'' is not visible, as it would do in the case of a very localized strong magnetic spot: a dipolar configuration naturally explains the observed behavior. However, even if the quadrupolar configuration is less favoured by our modeling procedure and if this topology does not appeal for an obvious formation scenario, we cannot rule it out as a possible solution.
As explained in Sect. 2.4, the present ZDI study was made using only the 20 spectra corresponding to about 3 rotations, between November 2007 and March 2010, to avoid epochs when activity appeared to deviate significantly form the mean variation. However, the main results as described above  would be the same if we employed the entire data set, except that the $\chi^2 _r$ would have been significantly worse.

From the dipolar model (corresponding to $i$ = 60$^\circ$ and Fig. 6) we can measure $\beta$ (the angle between rotational and magnetic axes of the dipole component) and the dipole strength  (since the dipolar component dominates, the strength of the dipole corresponds to about the maximum magnetic field measured on our map): we find $\beta$ = 45$^\circ$, $B_{\rm d} \sim$ 207 G. 

 We have also fit our data with a model with a pure dipole, corresponding to spherical harmonic $\ell = 1$ (and assuming $i$ = 60$^\circ$). 
The goodness of fit is (naturally) worse than for the models described above, and the inferred dipole parameters are $\beta$ = 22$^\circ$,  $B_{\rm d} \sim$  198 G.  

 The magnetic maximum is measured to occur at phases 0.04 and 0.96, respectively, for the two ZDI models described above.

It could also be of interest to use the measurements of $B_{\ell}$ from Sect. 2.2 (Table 1) to fit a dipolar oblique rotator model using the relations of Preston (1967) and reviewed by Auri\`ere et al. (2007). In this case we find  $\beta$ = 23$^\circ$ and dipole strength of 370 G (taking $i$ = 60$^\circ$ as suggested above). 

The smaller value obtained for  $B_{\rm d}$ from our ZDI models with respect to $B_{\rm d}$ inferred from the $B_{\ell}$ variations is due in part to the fact that our ZDI models correspond to Stokes $V$ signatures at phases near magnetic maximum that are weaker than those observed (Figure 5).

Taking $i$ = 60$^\circ$, in the case of rigid rotation, using $R$ = 4.68~R$_{\odot}$ from the evolutionary model of Charbonnel \& Lagarde (2010) used by Auri\`ere et al. (2008), we get $v\sin i$ = 0.69 km s$^{-1}$.

\begin{figure}
\centering
\includegraphics[width=7.5 cm,angle=0] {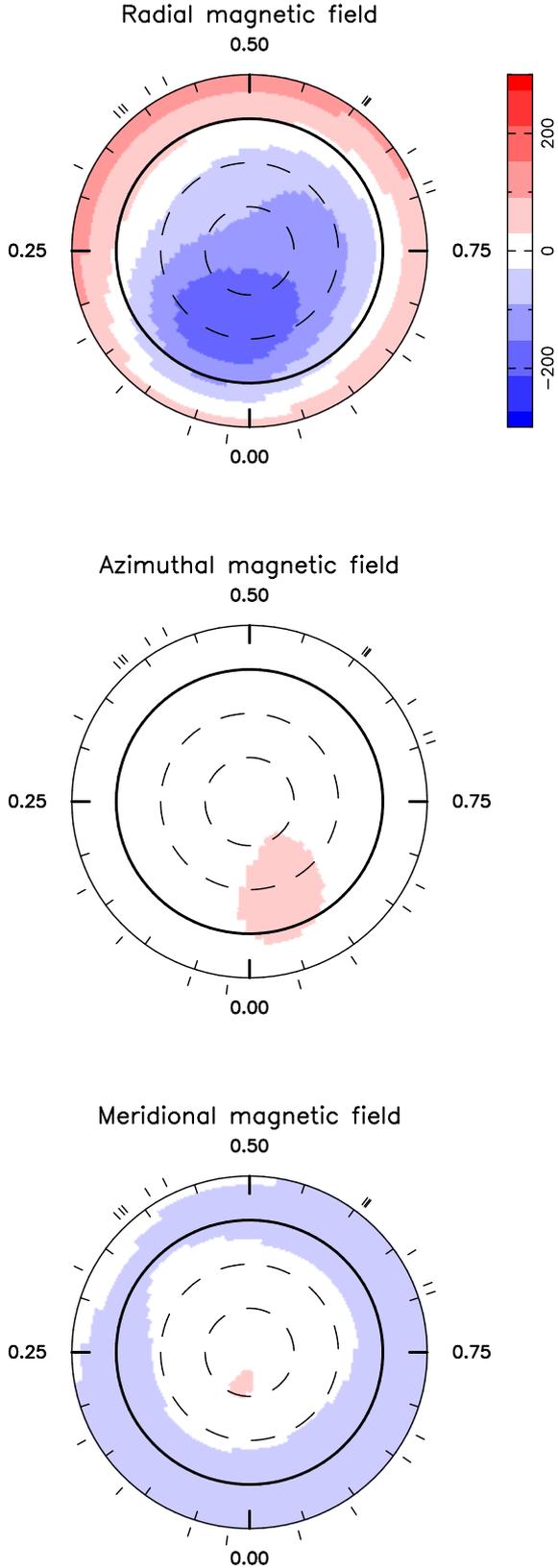} 
\caption{Magnetic map of EK Eri for $i = 60^\circ$. Each chart illustrates the field projection onto one axis of the spherical coordinate frame. The magnetic field strength is expressed in G. The star is shown in flattened polar projection down to latitudes of $-30^\circ$ , with the equator depicted as a bold circle and parallels as dashed circles. Radial ticks around each chart indicate the observed rotational phases.}
\end{figure} 
 
\section{Summary and discussion}

\subsection{Summary}

Spectropolarimetric observations of EK Eri have been obtained during 4 seasons, corresponding to about 4 photometric periods of 308.8 d. duration.

The longitudinal magnetic field $B_{\ell}$ is detected consistently, with $B_{\ell}$ variations of up to about 80 G without any reversal of its sign. The maximum of $|B_{\ell}|$ is observed to coincide with the phase of photometric light minimum (i.e. phase 0.0 of the photometric ephemeris of DBSS10).

The activity indicators measured from Ca~{\sc ii} H \& K, the Ca IR triplet and H$\alpha$ are shown to vary  with small amplitude and in phase with $|B_{\ell}|$. However some strong deviations are observed, which may suggest episodes of flares.

The radial velocity also varies in a reasonably coherent way according to the photometric ephemeris, but with a phase delay of about 0.1 cycles.

For our Zeeman Doppler imaging (ZDI) investigation we used the 308.8 d photometric period (DBSS10) which is similar to our best rotational period, based on quality-of-fit of the ZDI maps, of about 311 d.
 
For all the described  models we find that the large scale magnetic field of EK Eri is poloidal (more than 98 $\%$ of the magnetic energy is contained in this component). In addition, the surface magnetic field variations of EK Eri appear to be dominated by a strong magnetic spot (of negative polarity) which is phased with the darker, redder (and therefore presumably cool) photometric spot.

The ratio between dipolar and quadrupolar components of the magnetic field increases with decreasing value of the rotational axis inclination $i$. Whereas the quadrupolar component dominates for $i$ greater than 80$^\circ$, for $i$ = 60$^\circ$, we obtain a model that is almost purely dipolar. In the dipolar model, the photometric spot at phase 0 corresponds to the pole of negative polarity of the dipole, which could be the remnant of that of the Ap star progenitor of EK Eri.

\subsection{Discussion}

The evolution of a magnetic (2 $M_{\odot}$) Ap star from the main sequence to the red giant branch is characterized by a surface magnetic field which is expected to weaken (as $1/R^{2}$ if conservation of magnetic flux is assumed) while the convective envelope deepens. The rotational period of the surface also lengthens as the star expands. For an Ap star progenitor of EK Eri, one can infer a magnetic field of several thousand gauss and a very thin convective envelope on the main sequence (Stepie\'n 1993, Auri\`ere et al. 2008). At the main sequence evolutionary stage, the magnetic field would be sufficient to suppress convection (e.g. discussion by Th\'eado et al. 2005). At the present evolutionary stage of EK Eri we measure a large scale surface magnetic field dominated (for $i=60^\circ$) by a dipole of about 200 G strength, when the convective envelope of our 2 $M_{\odot}$ model contains $\sim 0.37$ $M_{\odot}$ of the star's mass (Auri\`ere et al. 2008). The outstanding magnetic activity properties of EK Eri are therefore the result of the interplay of the remaining magnetic field from one Ap star and deep convection. Such an interaction between a pre-existing magnetic field and thermal convection has been studied through numerical simulations mostly in the solar context (see e.g. Hurlburt et al. 1996, Cattaneo et al. 2003, Proctor 2004, Thomson 2005, Strugarek et al. 2011) and in the cases of core convection (Featherstone et al. 2009). For example, the transition between regimes when the imposed magnetic field is strong (magnetoconvection) and when the magnetic field is weak (dynamo) is investigated (e.g. Cattaneo et al. 2003).

No differential rotation is inferred from our modeling (Sect. 3.2): this can be expected both from a strong (fossil) magnetic field and from a slow rotator. Figure 5 shows that an average magnetic field can be defined during the 3 observed rotational periods used for modeling. Figure 2 shows that significant deviations occur during that period of time, and some are even stronger before (20 September 2007) and afterward (21 September and 13 October 2010; see also Sect. 2.4). This shows that even if  a magnetic dipole dominates the magnetic topology, we are not in the presence of a stable, large scale magnetic field as in an Ap star. Figure 5 also shows that it was not possible to perfectly fit our observations at phases near maximum magnetic strength. Therefore, the interplay of the remaining field and convection appears to induce some variable magnetic component. 

Another striking difference between the magnetic spot of EK Eri and the magnetic pole of an Ap star, is that for an Ap star surface features ("spots") correspond generally to chemical over-abundances of some elements, and are expected to be bright (Krticka et al. 2007,  L\"uftinger et al. 2010). In the case of EK Eri, the spot is dark and redder (and presumably cool), while the abundances are found to be similar to those of the sun at all phases (DBSS10; Van Eck et al. 2011 in preparation); the spot of EK Eri therefore appears to be a temperature spot, as observed in the late-type stars, where dynamo operates. As discussed by Auri\`ere et al. (2008) and Van Eck et al. (2011, in preparation), the overabundances observed at the surface of an Ap star would have been erased by the deep mixing occuring during the first dredge-up in which EK Eri is currently engaged. The differences between models for the magnetic poles of Ap stars and sunspots have been investigated (e.g. Shibahashi 2004).

EK Eri therefore appears to be an excellent laboratory to study the interplay between magnetic field and convection: the magnetic field at the surface of EK Eri may be considered as a new type of magnetic field, neither a rigid large scale magnetic field (e.g. dipole and oblique rotator model) nor a solar-type dynamo magnetic field. This magnetic field has the potential to strongly interact with the oscillations detected in EK Eri (Dall et al. DBSS10, 2011). However, the dipole field measured with our modeling from 2007-2010 data is rather weaker than the 300-500 G expected from preliminary computations (Dall et al. 2011) to be able to significantly alter the pulsations.

 This work shows that the magnetic variations of EK Eri are dominated by stellar rotation (i.e. rotational modulation of a dipole in our prefered scenario). The survival of a fossil dipole can explain the long term stability of the location of the dark spot, and magnetic cycles of varying photometric amplitude may be due to dynamo-like activity. Because of the very small $v\sin i$ of EK Eri, only large scale features of the surface magnetic field can be mapped with ZDI. However, the activity indicators are sensitive to all scales of magnetic regions. The observations of the 2010-2011 season show (Fig. 2) that while in September-October 2010 both the magnetic field and activity indicators were stronger than expected for that rotational phase, the magnetic field became ``normal'' at the beginning of 2011 while the values of the activity indicators remained high. Therefore the large scale magnetic component could be more stable than the small-scale field . In this context, it would be worthwhile to investigate if the possible magnetic cycles  suggested by photometry (DBSS10) exist as well in the magnetic data. Studying the correlation between photometric amplitude and surface magnetic strength/topology of EK Eri would certainly help to understand the interplay of the dipole remnant and convection.

\begin{acknowledgements}
       We thank  the TBL team for providing service observing with Narval and the PNPS of CNRS/INSU for financial support. Part of the observations in 2008 were funded under an OPTICON grant. R. K.-A. acknowledges the possibility of working for six months in 2010 as a visiting researcher in LATT, Tarbes, under Bulgarian NSF grant DSAB 02/3/2010, and partial financial support under NSF contract DO 02-85. C.C. acknowledges financial support from the Swiss National Science Foundation (FNS). GAW acknowledges Discovery Grant support from the Natural Science and Engineering Research Council of Canada.
\end{acknowledgements}


\begin{thebibliography} {}

\bibitem{} Auri\`ere M., Wade G.W., Silvester J., Ligni\`eres F. et al. 2007, A\&A, 475, 1053

\bibitem{} Auri\`ere M., Konstantinova-Antova R., Petit P. et al. 2008, A\&A, 499, 491

\bibitem{} Auri\`ere M., Wade G.A., Konstantinova-Antova R., et al. 2009, A\&A, 504, 231


\bibitem{} Cattaneo F., Emonet T., Weiss N. 2003, ApJ, 588, 1183

\bibitem{} Charbonnel C., Lagarde N. 2010, A\&A, 522, A10

\bibitem{} Dall T.H., Bruntt H., Strassmeier K.G., 2005, A\&A 444, 573

\bibitem{} Dall T.H., Bruntt H., Stello D., Strassmeier K.G. 2010, A\&A, 514, 25, (DBSS10)

\bibitem{} Dall T.H., Cunha M., Strassmeier et al. 2011, in Cool Star 16 meeting, (ASP Conferences Series), in press

\bibitem{} Donati J.-F. 2001, in Lecture Notes Phys. Vol. 573, Astrotomography: Indirect Imaging Methods in Observational Astronomy, ed. Boffin H., Steeghs D., Cuypers J. (Springer, Berlin), 207

\bibitem{} Donati J.-F., Semel M., Carter B.D. et al. 1997, MNRAS 291, 658

\bibitem{} Donati J.-F., Brown S. F., 1997, A\&A 326, 1135

\bibitem{} Donati J.-F., Catala C., Landstreet J., Petit P., 2006a, in Casini R., Lites B., eds, Solar Polarization Workshop n4, Vol.358 of ASPC series, 362

\bibitem{} Donati J.-F., Howarth I.D., Jardine M.M. et al., 2006b, MNRAS 370, 629

\bibitem{} Duncan D.K., Vaughan A.H., Wilson O. et al. 1991, ApJS, 76, 383

\bibitem{} Featherstone N.A., Browning M.K., Brun A.S., Toomre J. 2009, ApJ, 705, 1000

\bibitem{} Hurlburt N.E., Matthews P.C., Proctor M.R.E. 1996, ApJ, 457, 933

\bibitem{} Konstantinova-Antova R.K., Antov A.P. 2000, KFNTS, 3,342

\bibitem{} Konstantinova-Antova R.K., Antov A.P., Zhilyaev B.E. et al. 2005, AN, 326, 38

\bibitem{} Krticka J., Mikulasek Z., Zverko J., Ziznovsky J. 2007, A\&A, 470, 1089

\bibitem{} Lozitsky V.G., Baranovsky E.A., N.I. Lozitska N.I., Leiko U.M. 2000, Solar Phys., 191, 171

\bibitem{} L\"uftinger T., Fr\"ohlich H.-E., Weiss W.W. et al. 2010, A\&A, 509, 43

\bibitem{} Morgenthaler A., Petit P., Auri\`ere M. et al. 2011, in Cool Star 16 meeting, in press

\bibitem{} Morgenthaler A., Petit P., Saar S. et al. 2012, A\&A, submitted, 2011arXiv1109.5066M 

\bibitem{} Morin J., Donati J.-F., Petit P. et al. 2008, MNRAS, 390, 567

\bibitem{} Palacios A., Gebran M., Josselin E. et al. 2010, A\&A, 516, 13

\bibitem{} Petit P., Dintrans B., Solanki S.K. et al. 2008, MNRAS 388, 80

\bibitem{} Petit P., Donati J-F., Collier Cameron A. 2002, MNRAS, 334, 374

\bibitem{} Preston G.W. 1967, ApJ, 150, 547

\bibitem{} Proctor M. 2004, A\&G, 45, 4.14

\bibitem{} Rees D.E., Semel M., 1979, A\&A 74, 1

\bibitem{} Shibahashi H. 2004, in ``The A-Star Puzzle'', proceedings IAU Symposium $N_0$ 224, J. Zverko, J. Ziznovsky, S.J. Adelman, W.W. Weiss, eds. (Cambridge University Press, Cambridge), 451

\bibitem{} Stepie\'n K., 1993, ApJ 416, 368

\bibitem{} Stibbs D.W.N. 1950, MNRAS, 110, 395

\bibitem{}Strassmeier K.G., Stepie\'n K., Henry G.W., 1999, A\&A 343, 175

\bibitem{}Strugarek A., Brun A.S., Zahn J.-P. 2011, A\&A, 532, 34

\bibitem{} Th\'eado S., Vauclair S., Cunha M.S. 2005, A\&A, 443, 627

\bibitem{} Thomson S.D. 2005, MNRAS, 360, 1290

\bibitem{} Wright J.T., Marcy G. W., Butler R.P., Vogt S.S. 2004, ApJS, 152, 261

\bibitem{} Young A., Farding A., Thurman G., 1989, PASP 101, 1017


\end{thebibliography}
\end{document}